\documentclass[conference]{IEEEtran}
\IEEEoverridecommandlockouts
\usepackage{cite}
\usepackage{amsmath,amssymb,amsfonts}
\usepackage{algorithmic}
\usepackage{graphicx}
\usepackage{textcomp}
\usepackage{xcolor}
\usepackage{hyperref}
\usepackage{makecell}

\hypersetup{
    colorlinks=true, 
    linkcolor=black,  
    citecolor=black, 
    filecolor=black, 
    urlcolor=black    
}

\def\BibTeX{{\rm B\kern-.05em{\sc i\kern-.025em b}\kern-.08em
    T\kern-.1667em\lower.7ex\hbox{E}\kern-.125emX}}
\begin{document}

\title{Resource-Efficient Compilation of Distributed Quantum Circuits for Solving Large-Scale Wireless Communication Network Problems
\thanks{*Corresponding Author: kuan-cheng.chen17@imperial.ac.uk\\}
}

\author{
\IEEEauthorblockN{
    Kuan-Cheng Chen\IEEEauthorrefmark{2}\IEEEauthorrefmark{3}\IEEEauthorrefmark{1},
    Felix Burt\IEEEauthorrefmark{2}\IEEEauthorrefmark{3},
    Shang Yu \IEEEauthorrefmark{4}\IEEEauthorrefmark{3},
    Chen-Yu Liu\IEEEauthorrefmark{5}\IEEEauthorrefmark{6},
    Min-Hsiu Hsieh\IEEEauthorrefmark{6},
    Kin K. Leung \IEEEauthorrefmark{2}\IEEEauthorrefmark{3}
}
\IEEEauthorblockA{\IEEEauthorrefmark{2}Department of Electrical and Electronic Engineering, Imperial College London, London, UK}
\IEEEauthorblockA{\IEEEauthorrefmark{3}Centre for Quantum Engineering, Science and Technology (QuEST), Imperial College London, London, UK}
\IEEEauthorblockA{\IEEEauthorrefmark{4}Department of Physics, Imperial College London, London, UK}
\IEEEauthorblockA{\IEEEauthorrefmark{5}Graduate Institute of Applied Physics, National Taiwan University, Taipei, Taiwan}
\IEEEauthorblockA{\IEEEauthorrefmark{6}Hon Hai Research Institute, Taipei, Taiwan}
}

\maketitle

\begin{abstract}

Optimizing routing in Wireless Sensor Networks (WSNs) is pivotal for minimizing energy consumption and extending network lifetime. This paper introduces a resource-efficient compilation method for distributed quantum circuits tailored to address large-scale WSN routing problems. Leveraging a hybrid classical-quantum framework, we employ spectral clustering for network partitioning and the Quantum Approximate Optimization Algorithm (QAOA) for optimizing routing within manageable subgraphs. We formulate the routing problem as a Quadratic Unconstrained Binary Optimization (QUBO) problem, providing comprehensive mathematical formulations and complexity analyses. Comparative evaluations against traditional classical algorithms demonstrate significant energy savings and enhanced scalability. Our approach underscores the potential of integrating quantum computing techniques into wireless communication networks, offering a scalable and efficient solution for future network optimization challenges.

\end{abstract}


\section{Introduction}

\begin{figure}[!t]
\centering
\includegraphics[scale=0.45]{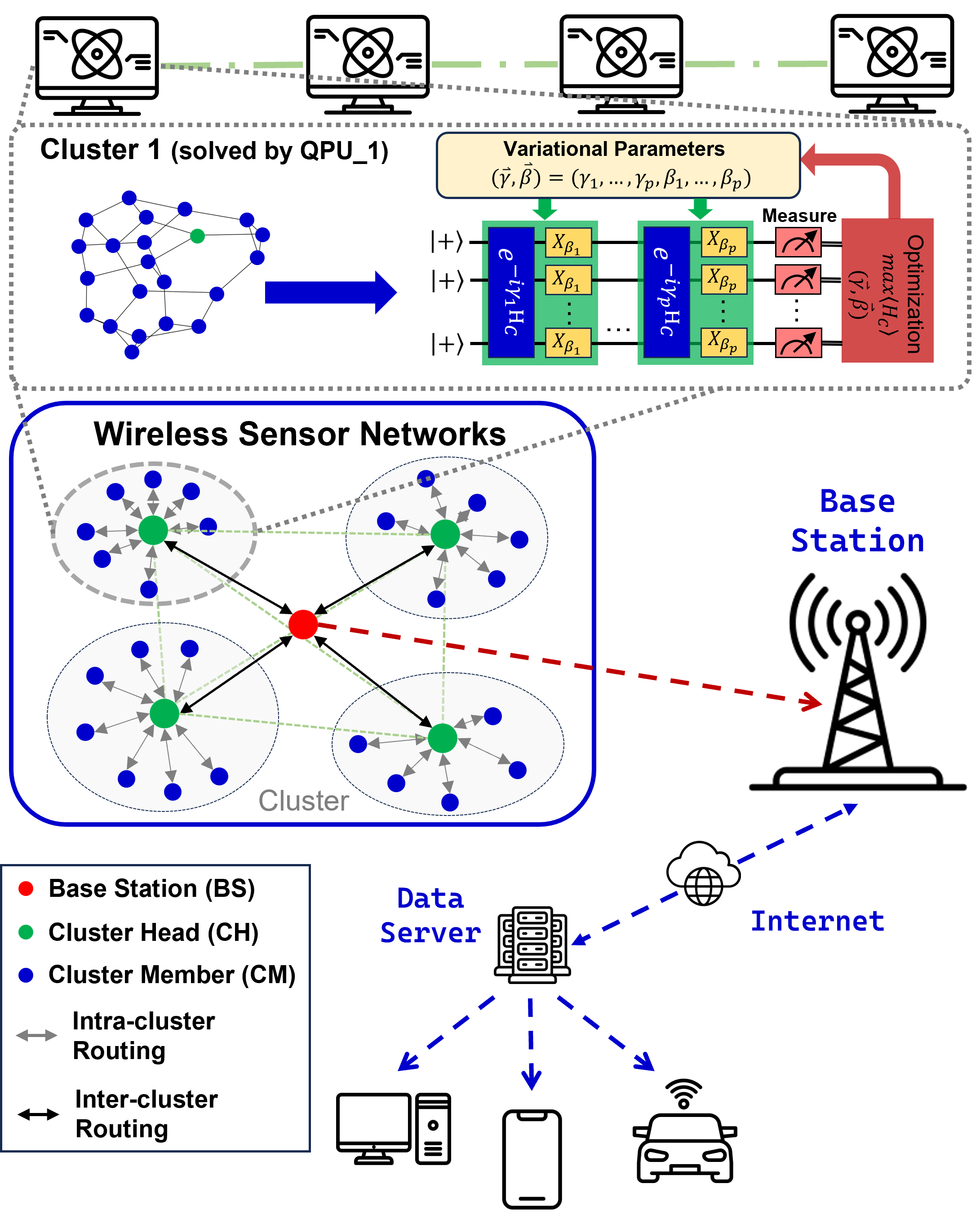}
\caption{Conceptual overview of the hybrid classical-quantum approach for optimizing WSN routing. The network is partitioned into clusters, each treated as a subgraph optimized by QPUs using the QAOA. The quantum circuit solves intra-cluster routing, while classical methods handle inter-cluster routing. The network comprises base stations (BS, red), cluster heads (CH, green), and cluster members (CM, blue). Detailed optimization methods and routing strategies are further explained later in the paper.
 }
\label{fig:concept}
\end{figure}

The rapid evolution of communication networks and distributed systems has led to increasingly complex optimization challenges, particularly in large-scale, data-intensive environments\cite{you2021towards,alsabah20216g,cai2022compute}. From traffic management in next-generation wireless networks to resource allocation in cloud computing and Internet of Things (IoT) ecosystems, the need for efficient optimization techniques is more critical than ever\cite{djahel2014communications,sheng2013survey}. As these systems expand, conventional optimization algorithms\cite{liu2024survey,pham2020whale} face scalability limits due to the sheer size of the problem space, complexity constraints, and real-time processing requirements. In this context, the emergence of quantum computing offers a promising frontier to tackle such challenges with enhanced computational capabilities\cite{moll2018quantum}.

The Quantum Approximate Optimization Algorithm (QAOA) has gained significant attention for its ability to solve combinatorial optimization problems more efficiently than classical methods\cite{farhi2014quantum}. By leveraging quantum devices to explore multiple solution pathways simultaneously, QAOA is well-suited for problems like traffic routing\cite{azad2022solving}, resource scheduling\cite{amaro2022case}, and clustering\cite{moussa2022unsupervised}. The hybrid nature of QAOA combines quantum circuits for generating candidate solutions with classical algorithms for refining these results, offering a powerful framework for tackling large-scale, NP-hard network optimization problems\cite{chen2024noise, zhou2023qaoa}. However, the practical application of QAOA is constrained by current quantum hardware limitations, including qubit coherence, gate fidelity, and restricted qubit connectivity, making direct quantum processing of large-scale problems impractical\cite{zhou2020quantum,blekos2024review}. To address this, hybrid classical-quantum frameworks have emerged, where the problem is divided into smaller subproblems solvable by distributed quantum processing units (QPUs), while classical methods manage inter-cluster optimization and refinement, overcoming hardware constraints while maintaining efficiency\cite{chen2024noise}.

A relevant example demonstrating the potential of hybrid QAOA is Wireless Sensor Networks (WSNs)\cite{raghavendra2006wireless}, which consist of numerous sensor nodes communicating to perform distributed tasks. WSNs face combinatorial optimization challenges, particularly in routing and energy efficiency, where the primary goal is to minimize energy consumption and extend the network's operational lifetime by optimizing communication paths\cite{kocakulak2017overview}. These problems are NP-hard\cite{nguyen2018mobile}, and as the scale of WSNs increases, classical algorithms struggle to efficiently find optimal solutions. While current QAOA implementations lack qubit fidelity (require quantum error correction\cite{he2024performance}) and scalability to handle large-scale optimization directly, WSNs serve as an ideal testbed for hybrid QAOA approaches. By partitioning the network into smaller clusters and optimizing each subproblem on quantum hardware, hybrid methods offer a feasible and scalable solution to complex network optimization tasks, illustrating the practical utility of quantum-classical frameworks.

This paper introduces a resource-efficient compilation framework for distributed quantum circuits (particularly IBM Quantum's heavy hexagonal devices\cite{gambetta2020ibm}), designed to address large-scale optimization problems in modern communication networks, with WSNs as a key use case. By partitioning the network into smaller clusters and applying QAOA to subgraphs, we mitigate the scalability issues associated with current quantum hardware. The framework efficiently leverages quantum resources while ensuring that network-level performance metrics such as energy consumption, latency, and throughput are optimized.

\section{Methodology}
\subsection{Network Modeling}

We model the WSN as a directed graph \( G = (V, E) \), where \( V \) represents the set of nodes and \( E \) denotes the set of edges corresponding to communication links between nodes. The node set \( V \) comprises sensor nodes (\( S \)), CHs (\( C \)), and the base station (\( B \)). Each node \( i \in V \) is characterized by its role \( r_i \) and initial energy \( E_i \). The role \( r_i \) is defined as \( r_i \in \{ \text{Sensor}, \text{CH}, \text{BS} \} \), indicating whether the node is a sensor, a cluster head (CH), or the base station (BS). The initial energy \( E_i \) is assigned based on the node's role\cite{heinzelman2000energy}:

\[
E_i = 
\begin{cases}
100 & \text{if } r_i = \text{Sensor}, \\
200 & \text{if } r_i = \text{CH}, \\
\infty & \text{if } r_i = \text{BS}.
\end{cases}
\]

\subsection{Communication Range and Edge Establishment}

Nodes communicate if they are within a predefined communication range \( R \). The Euclidean distance between nodes \( i \) and \( j \) is calculated as:

\begin{equation}
d_{ij} = \sqrt{(x_i - x_j)^2 + (y_i - y_j)^2},
\end{equation}

where \( (x_i, y_i) \) and \( (x_j, y_j) \) are the coordinates of nodes \( i \) and \( j \), respectively. An edge \( (i, j) \in E \) is established if \( d_{ij} \leq R \). The energy cost \( c_{ij} \) for transmitting data from node \( i \) to node \( j \) follows the free-space path loss model \cite{rappaport2024wireless}:

\begin{equation}
c_{ij} = \varepsilon \cdot d_{ij}^2,
\end{equation}

where \( \varepsilon \) is the energy consumption coefficient.

\subsection{Spectral Clustering for Network Partitioning}

To decompose the network into manageable subgraphs, we employ spectral clustering \cite{luxburg2007tutorial}. This method partitions the network into \( k \) clusters, where \( k \) is the number of CHs. Initially, an adjacency matrix \( A \) is constructed using the \( k \)-nearest neighbors approach, defined as:

\[
A_{ij} = 
\begin{cases} 
1 & \text{if node } j \text{ is among the } k \text{ nearest neighbors of node } i, \\
0 & \text{otherwise}.
\end{cases}
\]

Subsequently, the Laplacian matrix \( L \) is computed as:

\[
L = D - A,
\]

where \( D \) is the degree matrix with diagonal entries \( D_{ii} = \sum_{j} A_{ij} \). Eigenvalue decomposition is then performed on \( L \) to obtain the first \( k \) eigenvectors corresponding to the smallest \( k \) eigenvalues, capturing the essential structure of the graph. K-means clustering is applied to the rows of the matrix formed by these eigenvectors, effectively partitioning the nodes into \( k \) clusters. Each cluster corresponds to a subgraph \( G_s = (V_s, E_s) \), where \( V_s \subset V \) and \( E_s \subset E \).

\subsection{Formulating the QUBO Problem}

For each subgraph \( G_s \), the routing optimization is formulated as a Quadratic Unconstrained Binary Optimization (QUBO) problem\cite{lewis2017quadratic}. The objective is to minimize the total energy consumption while satisfying flow conservation and energy constraints. Binary variables \( x_{ij} \in \{0, 1\} \) are defined for all \( (i, j) \in E_s \), where \( x_{ij} = 1 \) indicates that the edge \( (i, j) \) is included in the routing path.

The QUBO objective function is expressed as:

\begin{align}
\text{Minimize } \quad & \sum_{(i,j) \in E_s} c_{ij} x_{ij} \notag \\
& + \lambda_{\text{flow}} \cdot \text{FlowConstraints}(\mathbf{x}) \notag \\
& + \lambda_{\text{energy}} \cdot \text{EnergyConstraints}(\mathbf{x}),
\end{align}

where \( \lambda_{\text{flow}} \) and \( \lambda_{\text{energy}} \) are penalty coefficients for the flow conservation and energy constraints, respectively.

To ensure flow conservation, the following constraint is imposed for each node \( i \in V_s \):

\begin{equation}
\sum_{j: (i,j) \in E_s} x_{ij} - \sum_{j: (j,i) \in E_s} x_{ji} = b_i,
\end{equation}

where \( b_i \) is the net flow at node \( i \):

\[
b_i =
\begin{cases}
1 & \text{if } r_i = \text{Sensor}, \\
0 & \text{if } r_i = \text{CH}, \\
- \sum_{i \in S} b_i & \text{if } r_i = \text{BS}.
\end{cases}
\]

Energy constraints are enforced to ensure that the energy consumed by a node does not exceed its initial energy \( E_i \):

\begin{equation}
\sum_{j: (i,j) \in E_s} c_{ij} x_{ij} \leq E_i.
\end{equation}

These constraints are incorporated into the QUBO objective using penalty terms, ensuring that feasible solutions satisfy both flow conservation and energy limitations.

\subsection{Resource-Efficient Distributed QAOA}

We utilize the QAOA to solve the QUBO problem for each subgraph \( G_s \) \cite{farhi2014quantum,hadfield2019from}. QAOA constructs a quantum state \( |\psi(\gamma, \beta)\rangle \) through the application of alternating problem-specific and mixing Hamiltonians, formulated as:
\begin{equation} 
|\psi(\gamma, \beta)\rangle = U_M(\beta_p) U_P(\gamma_p) \cdots U_M(\beta_1) U_P(\gamma_1) |s\rangle, 
\end{equation}

where \( U_P(\gamma) = \exp(-i \gamma H_P) \) encodes the QUBO objective as the problem Hamiltonian unitary operator, \( U_M(\beta) = \exp(-i \beta H_M) \) represents the mixing Hamiltonian unitary operator, \( |s\rangle \) is the initial uniform superposition state, and \( p \) denotes the number of layers. The parameters \( \gamma \) and \( \beta \) are optimized using classical methods to minimize the expected value \( \langle \psi(\gamma, \beta) | H_P | \psi(\gamma, \beta) \rangle \). This iterative process applies alternating quantum operations followed by classical parameter updates to converge on optimal solutions.

\begin{figure}[!b]
    \centering
    \includegraphics[width=1\linewidth]{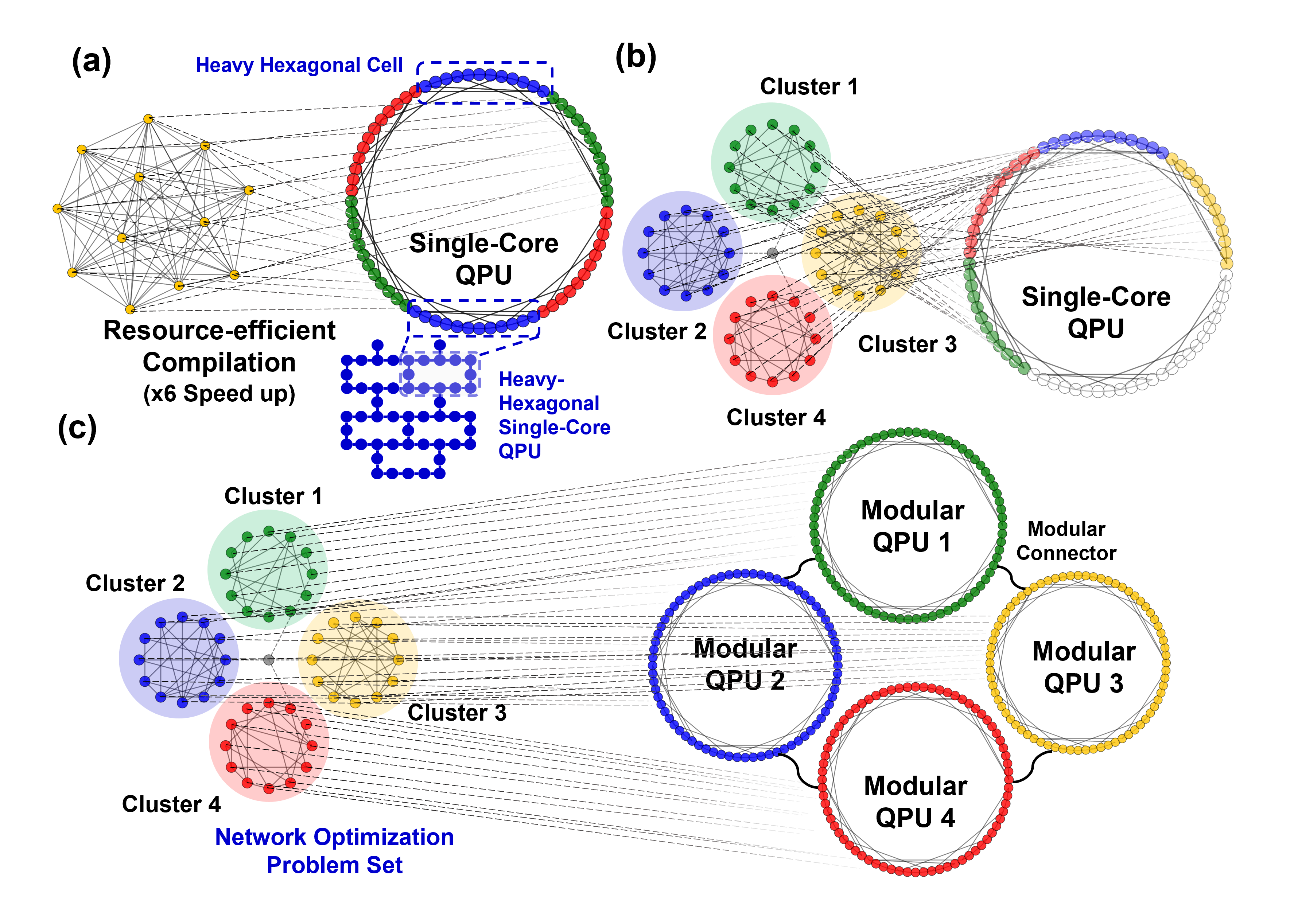}
    \caption{. Compilation strategies for resource-efficient distributed QAOA. (a) Resource-efficient compilation strategy for a single-core heavy-hexagonal QPU\cite{gambetta2020ibm}. (b) Cluster-based compilation strategy, where four distinct clusters are compiled and connected to a shared QPU. (c) Modular distributed compilation strategy, involving four independent QPUs connected through modular connectors, supporting efficient inter-cluster communication and execution.}
    \label{fig:QAOA_Compilation}
\end{figure}

The performance of QAOA is determined by the quantum circuit depth, which depends on the number of variables \( n_s \) and the number of layers \( p \), as well as the complexity of the classical optimization process, which scales with the number of parameters \( 2p \). To accommodate hardware limitations, subgraphs are constrained to a maximum size \( n_s \leq n_{\max} \). Larger subgraphs that exceed this limit are processed using classical methods, ensuring scalability and adaptability to real-world hardware constraints. This approach is illustrated in Fig. \ref{fig:QAOA_Compilation}, where a resource-efficient compilation strategy is deployed across various QPUs. The system begins with a single-core QPU optimized with a heavy-hexagonal architecture, achieving linear speedup for sampling. It then extends to distributed execution by clustering qubits within the QPU, and further scales to a modular QPU setup with interconnected units, allowing for efficient handling of large-scale network optimization problems while leveraging the benefits of distributed QAOA\cite{chen2024noise}.

\subsection{Network-Level Optimization and Performance Metrics}

After optimizing each subgraph, ensuring full network connectivity is crucial. This process involves several steps. Firstly, eachCH must be connected to the BS or another CH. If the distance \( d_{i0} \) between CH \( i \) and the BS satisfies \( d_{i0} \leq R \), an edge \( (i, 0) \) is established. Otherwise, the CH connects to the nearest CH within the communication range. Subsequently, a breadth-first search (BFS) is performed starting from the BS to identify any disconnected nodes. For any such disconnected nodes, edges are added to the nearest connected nodes in a manner that minimizes additional energy costs. This comprehensive approach ensures that the entire network remains connected and functional.

To evaluate the effectiveness of the optimization, several performance metrics are calculated\cite{lee2014modeling}. The total energy consumption \( C_{\text{total}} \) is defined as:

\begin{equation}
C_{\text{total}} = \sum_{(i,j) \in E_{\text{opt}}} c_{ij} x_{ij},
\end{equation}

where \( E_{\text{opt}} \) denotes the set of edges in the optimized network. Energy savings \( \Delta C \) are determined by:

\begin{equation}
\Delta C = C_{\text{initial}} - C_{\text{total}},
\end{equation}

with \( C_{\text{initial}} \) representing the total energy consumption of the initial network. The percentage reduction in energy consumption is then calculated as:

\begin{equation}
\text{Reduction (\%)} = \left( \frac{\Delta C}{C_{\text{initial}}} \right) \times 100\%.
\end{equation}

These metrics provide quantitative measures of the optimization's impact on energy efficiency.

\section{Results}
We simulated a WSN comprising 109 nodes, which included 100 sensor nodes, 8 CHs, and 1 BS. The nodes were randomly positioned within a \(100 \times 100\) unit area, with the BS strategically located at \((50, 110)\) to facilitate optimal network coverage. The communication range was set to \( R = 25 \) units, ensuring that nodes could establish reliable communication links within this radius.

\begin{figure}[htbp]
    \centering
    \includegraphics[width=0.95\linewidth]{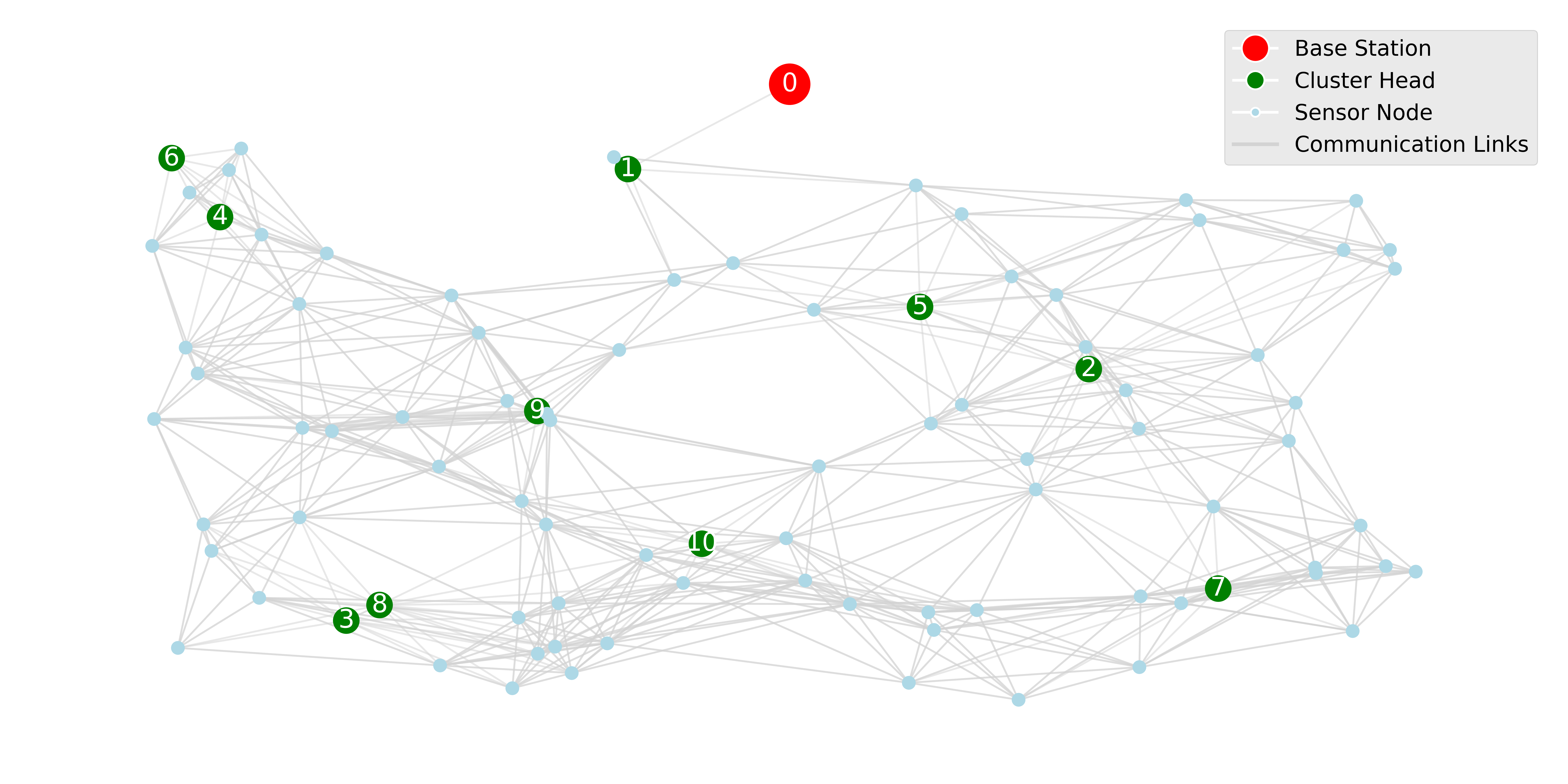}
    \caption{Initial Wireless Sensor Network Topology with Clusters Highlighted.}
    \label{fig:initial_network}
\end{figure}

To effectively manage the network's scalability and complexity, we employed spectral clustering to partition the network into 5 distinct clusters, each corresponding to a CH. This partitioning resulted in the formation of subgraphs \( G_s = (V_s, E_s) \), where each subgraph represents a manageable segment of the overall network. Fig.~\ref{fig:initial_network} illustrates the initial network topology with clusters distinctly highlighted, providing a visual representation of the clustered network structure.

For each subgraph \( G_s \), we formulated the routing optimization problem as a QUBO model and applied the QAOA to determine the optimal routing paths. Due to the limitations of current quantum hardware in the NISQ era\cite{preskill2018quantum}, classical hardware was employed to simulate the QAOA quantum circuit\cite{xu2024hamiltoniq,lykov2023fast}. However, to ensure computational feasibility given the limitations of simulating quantum circuits \cite{zhou2020limits}, subgraphs were restricted to a maximum of  \( n_s \leq 25 \) variables. The QUBO formulation incorporated energy consumption, flow conservation, and energy constraints, as detailed below:
\begin{align}
&\text{Minimize } \quad  \sum_{(i,j) \in E_s} c_{ij} x_{ij} \notag \\
& + \lambda_{\text{flow}} \sum_{i \in V_s} \left( 
    \sum_{j: (i,j) \in E_s} x_{ij} 
    - \sum_{j: (j,i) \in E_s} x_{ji} 
    - b_i 
\right)^2 \notag \\
& + \lambda_{\text{energy}} \sum_{i \in V_s} \left( 
    \sum_{j: (i,j) \in E_s} c_{ij} x_{ij} 
    - E_i 
\right)^2.
\end{align}

\begin{figure}[htbp]
    \centering
    \includegraphics[width=0.95\linewidth]{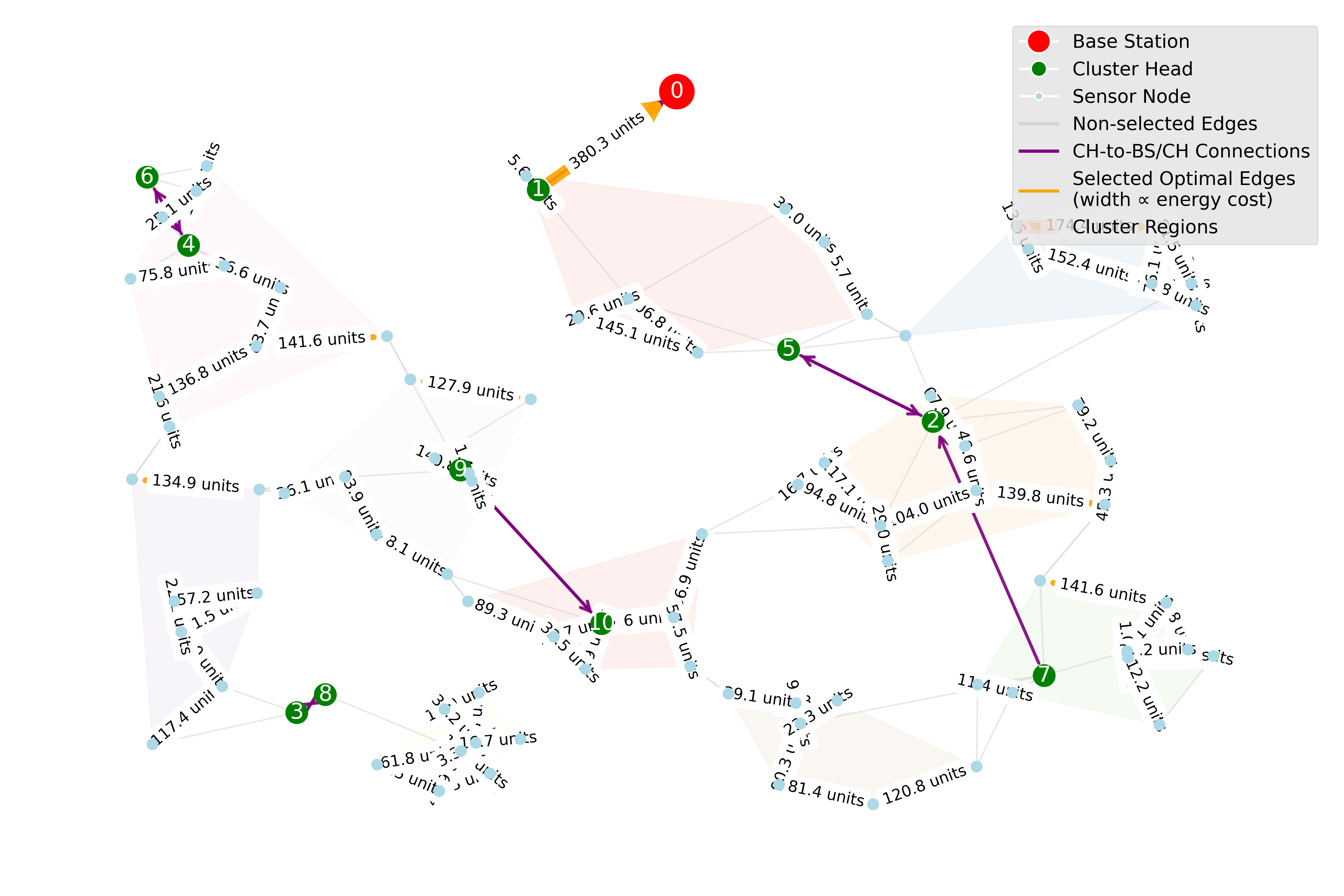}
    \caption{Optimized Routing Paths by Hybrid QAOA Approach.}
    \label{fig:optimized_network}
\end{figure}

For proof of concept, the implementation of our hybrid QAOA approach yielded better energy savings compared to the classical method, which employed a greedy search in subgroups. The initial energy cost of the network was calculated to be \( C_{\text{initial}} = 94,593.5 \) units. Using the greedy search approach, the total energy consumption reduced to \( C_{\text{total, classical}} = 29,969.7 \) units, resulting in energy savings of \( \Delta C_{\text{classical}} = 64,623.8 \) units, corresponding to a percentage reduction of \( 68.32\% \). In contrast, the quantum-enhanced optimization further reduced the total energy consumption to \( C_{\text{total, quantum}} = 15,982.9 \) units, yielding energy savings of \( \Delta C_{\text{quantum}} = 78,901.6 \) units, which represents a larger percentage reduction of \( 83.16\% \). This comparison highlights the superior performance of quantum-assisted optimization in minimizing energy consumption, surpassing the results achieved by the classical greedy search method in subgroups. Fig.~\ref{fig:optimized_network} illustrates the optimized network topology for both methods, with the selected edges highlighted in proportion to their energy costs.


\section{Conclusion}
This study introduces a hybrid classical-quantum framework for optimizing WSNs, offering notable advantages in scalability, energy efficiency, and resource utilization. By partitioning the network into smaller subgraphs, the framework overcomes the limitations of current quantum hardware, facilitating the optimization of large-scale networks. The application of the QAOA leads to substantial reductions in total energy consumption. Additionally, by constraining subgraph sizes to a maximum threshold, the approach ensures feasibility with existing quantum resources while leveraging classical processing for managing larger network scales. The computational complexity of the proposed method is influenced by both classical and quantum components, with spectral clustering exhibiting a complexity of \( O(N^3) \) \cite{ding2024survey} and QAOA's circuit depth scaling as \( O(p \times \log(n_s)) \)\cite{blekos2024review}, where \( N \) is the number of nodes and \( p \) the number of QAOA layers. Despite these strengths, the framework faces limitations, including constraints imposed by current quantum hardware, the dependency of QAOA on parameter selection which may affect solution optimality, and the scalability challenges associated with classical spectral clustering in extremely large networks. Future work will focus on enhancing the algorithmic aspects by exploring error mitigation and error correction with distributed QAOA on real quantum hardware\cite{he2024performance, chen2023short,prest2023quantum}, as well as optimizing compilation strategies leveraging E-bits\cite{burt2024generalised}.

\clearpage

\bibliographystyle{ieeetr}
\bibliography{references}

\end{document}